\documentclass[a4paper,twoside]{article}

\usepackage[T1]{fontenc}
\usepackage[utf8]{inputenc}

\usepackage{lmodern}		

\usepackage[british]{babel}
\usepackage{microtype}

\usepackage{amssymb}	
\usepackage{amsmath}	
\usepackage{amsthm}
\usepackage{bbm}
\usepackage{bm}
\usepackage{mathrsfs}
\usepackage{mathtools}
\usepackage{braket}
\usepackage{dsfont}

\usepackage{siunitx}
\usepackage{textcomp}

\usepackage{booktabs}
\usepackage{graphicx}
\usepackage{multirow}
\usepackage{tabularx}
\usepackage[labelfont={bf},font=small]{caption}
\usepackage{subfig}
\usepackage{dcolumn}
\newcolumntype{d}[1]{D..{#1}}

\usepackage{enumerate}		
\usepackage[italian]{varioref}	
\usepackage{indentfirst}	
\usepackage{emptypage}		
\usepackage{lipsum}			
\usepackage{comment}		
\usepackage{syntonly}		
\usepackage{quoting}		
\usepackage{multicol}		

\usepackage[babel]{csquotes}
\usepackage[style=authoryear]{biblatex}
\bibliography{biblio}

\DeclareMathOperator{\cov}{cov}
\DeclareMathOperator{\define}{\equiv}
\DeclareMathOperator{\E}{\mathbb{E}}

\DeclareMathOperator{\Lik}{\mathcal{L}}

\DeclareMathOperator{\norm}{\mathcal{N}}
\DeclareMathOperator{\Real}{\mathbb{R}}
\DeclareMathOperator{\trace}{tr}

\DeclarePairedDelimiter{\norma}{\lVert}{\rVert}

\newcommand{\diff}{\text{d}\kern-.05em}
\newcommand{\inv}{^{-1}}
\renewcommand{\Pr}{\mathbb{P}}
\newcommand{\T}{^\mathrm{\scriptscriptstyle T}}

\newcommand{\ind}[1]{\mathds{1}_{\{#1\}}}

\title{Reduced-bias estimation of spatial econometric models with incompletely geocoded data}
\author{Giuseppe Arbia, Maria Michela Dickson\\ Giuseppe Espa, Diego Giuliani, Flavio Santi}

\begin{document}

\maketitle

\begin{abstract}
The application of state-of-the-art spatial econometric models requires that the information about the spatial coordinates of statistical units is completely accurate, which is usually the case in the context of areal data. With micro-geographic point-level data, however, such information is inevitably affected by locational errors, that can be generated intentionally by the data producer for privacy protection or can be due to inaccuracy of the geocoding procedures. This unfortunate circumstance can potentially limit the use of the spatial econometric modelling framework for the analysis of micro data. Indeed, some recent contributions~\parencite[see e.g][]{arbia2016} have shown that the presence of locational errors may have a non-negligible impact on the results. In particular, wrong spatial coordinates can lead to downward bias and increased variance in the estimation of model parameters. 

This contribution aims at developing a strategy to reduce the bias and produce more reliable inference for spatial econometrics models with location errors. The validity of the proposed approach is assessed by means of a Monte Carlo simulation study under different real-case scenarios. The study results show that the method is promising and can make the spatial econometric modelling of micro-geographic data possible.
\end{abstract}

\section{Introduction}
\label{sec:intro}

Traditional spatial econometric models are based on the implicit assumption that the information about the spatial location of statistical units is completely accurate. Whilst this circumstance is the norm in the context of areal data (such as municipalities, counties or regions), it is rarely met when the observations are points in space (such as firms, houses or facilities), whose locations may be either missing or affected by locational errors (see \cite{zimmerman2008,zimmerman2010,arbia2019a}).

Although geolocation may fail for some units because of technical reasons, incomplete positioning arises more frequently in geocoding processes, especially in those circumstances where units' coordinates are obtained by matching units' postal addresses with georeferenced street maps~(see e.g. \cite{kravets2007}). Clearly, the quality of the resulting geolocation depends both on the correctness and completeness of postal addresses, as well as on the effectiveness of matching algorithms and softwares, nonetheless, if position of some units is uncertain, this fact should be properly considered in the estimation process. 

When an incomplete address is geocoded, unit's position is conventionally imputed to the centroid of the area where unit is located, as it can be known from address information. Such areas may be counties, municipalities, or, more frequently, ZIP code areas~\cite{zimmerman2008}. From a statistical point of view, the presence of locational errors due to coarsened locations may have a significant impact on parameter estimates of spatial econometric models based on the Cliff-Ord approach~\cite{cliff1969}, as positional errors lead to downward biased estimates for the spatial autoregressive parameters and inconsistent estimates for covariates coefficients \cite{arbia2016}.

This paper tackles the problem of estimating spatial models where part of units are affected by coarsening. In particular, we focus on the Spatial Lag Model~(see e.g, \cite{arbia2014}). The proposed estimation strategy models both the spatial stochastic process and the coarsening mechanism by means of a marked point process whose intensity function is estimated according to the coarsened-data estimator proposed by \cite{zimmerman2008}. Model is fitted through the maximisation of a doubly-marginalised likelihood function of the marked point process, which cleans out the effects of coarsening.

The first marginalisation of the likelihood function allows the dimensionality of the spatial econometric model to be consistently reduced to non-coarsened points and it is derived analytically. The second marginalisation is performed via Monte Carlo simulations over the locations of coarsened points.

The modelling approach and Monte Carlo experiments presented in the paper show the validity of the proposed estimation method in comparison with the estimates obtained by means of other estimation approaches. In particular, the comparison concerns the parameter estimates and the direct and indirect effects of model covariates on the dependent variable \cite{arbia2019b}.

The paper is organised as follows. Section~\ref{sec:model} describes the modelling approach and the notation we adopted in this paper. Section~\ref{sec:estimation} illustrates and discusses the proposed estimation approach. Section~\ref{sec:simulations} illustrates the results of Monte Carlo simulations where the finite properties of parameters' estimators and direct and of indirect impacts of regressors are studied. Section~\ref{sec:conclusions} concludes the paper.

\section{Modelling approach and notation}
\label{sec:model}

Consider a population of $n$ units $i=1,\dots,n$ for which a quantitative characteristic of interest $y_i\in\Real$ and $k$ regressors $x_i\in\Real^k$ are known. Assume that postal addresses are available for all $n$ units, however only $p<n$ of them are complete, whereas $n-p$ are incomplete. Assume also, that the $p$ units can be assigned to, say, the ZIP areas they actually belong to.

Under these conditions, if a spatial model is needed for modelling $y$~(a thorough illustration of the reasons why a spatial modelling approach may be necessary is available in \cite[ch. 2]{lesage2009}), the coarsening of the $n-p$ units' locations only affects the specification of the spatial weight matrix, as $y_i$ and $x_i$ are known for all units $i=1,\dots,n$.

Consider, for example, the following isotropic Spatial Lag Model (SLM):
\begin{equation}
\label{eq:SLM}
\begin{cases}
y=\rho Wy+X\beta+\varepsilon\\
\varepsilon\sim\norm_n(0,\sigma^2I_n)
\end{cases}
\end{equation}
\noindent where $X\in\Real^{n\times k}$ is the design matrix which includes $k$ regressors, and $W\in\Real^{n\times n}$ is the usual spatial weight matrix whose elements $w_{ij}$ take positive values according to some proximity criterion and zero if units $i$ and $j$ are not considered as neighbours.

It can be verified that, if $p/n$ is the proportion of non-coarsened units, the share of elements of $W$ not affected by coarsening is only about $(p/n)^2$, whereas all elements change if $W$ is stochastic (that is, if $W$ is row-standardised). The magnitude of the effects of coarsening on the spatial weight matrix is the cause of bias of estimators for the autoregressive parameter $\rho$ \cite{arbia2016}. 

The estimation method proposed in this paper basically reduces the dimensionality of the model by concentrating the likelihood on the $p$ non-coarsened units, thus limiting the effects of the coarsened locations on model estimates and, at the same time, exploiting the available information about covariates and zone-based location of the coarsened units.

The problem is modelled as a marked point process where both the stochastic spatial process and the coarsening process are specified conditionally on the underlying point process.

Let $(\Omega, \mathcal{F}, \Pr)$ be a probability space, and let $Z\in\Real^{n\times2}$ be a realisation of $n$ points from a 2-dimensional point process $\set{Z(s,\omega)\colon s\in S}$ defined over
a bounded metric space $(S,\norma{\cdot})$ where $S\subset\Real^2$. Let $\lambda\colon S\to\Real^+$ be the intensity function of $\set{Z(s,\omega)\colon s\in S}$ defined as:
\[
\lambda(x)=\lim_{|\diff x|\to0}\frac{\E(N(x,\diff x))}{\diff x}\,,
\]
being $N(x,\diff x)$ the count function for points in the neighbour $\diff x\subset S$ centered in $x\in S$ (see e.g. \cite{illian2008}).

Conditionally on $Z$, the isotropic SLM~\eqref{eq:SLM} is defined for the spatial process $y$, where the spatial weight matrix $W$ is row-standardised and its elements $w_{ij}$ are defined as follows:
\begin{equation}
\label{eq:wfun}
w_{ij}=
\begin{cases}
\frac{\kappa(\norma{z_i-z_j})}{\sum_{h=1}^n\kappa(\norma{z_i-z_h})} & \text{if $i\neq j$ and $\sum_{h=1}^n\kappa(\norma{z_i-z_h})\neq0$}\\
0 & \text{otherwise}\\
\end{cases}\,,
\end{equation}
for any $i,j\in\{1,\dots,n\}$, and some non-increasing function $\kappa\colon\Real^+\to\Real^+$ such that $\lim_{x\to\infty}\kappa(x)=0$.

The coarsening process can be either dependent on the intensity function $\lambda$ and the realisation of the point process $\set{Z(s,\omega)\colon s\in S}$ or independent from them. Here we just assume that the coarsening is modelled by means of a random vector $\Phi$, which is a realisation of $n$ Bernoulli random variables independent from the spatial process $y$ conditionally on the point process $Z$. The components $\Phi_j$ of the random vector $\Phi$ are defined as follows:
\begin{equation}
\label{eq:coarseningprocess}
\Phi_j\sim\mathcal{B}(p_j)
\end{equation}
for $j=1,\dots,n$, and take value $\Phi_j=0$ if point $j$ is coarsened, whereas $\Phi_j=1$ if point $j$ has been correctly geocoded.

Finally, let $\mathcal{S}=\set{S_1,S_2,\dots,S_R}$ be a partition of the space $S$ into $R$ regions such that, for any unit $i$ with coordinate $c_i\in S$, it exists one region $S_r$ such that $c_i\in S_r$.\footnote{In fact, this assumption is not crucial in our analysis, and can be easily generalised by assuming $\mathcal{S}$ to be a cover of $S$ such that $S\in\mathcal{S}$. This generalisation permits various degrees of incompleteness in postal addresses to be modelled, including the situation where some units are only known to be located in $S$. The estimation method proposed later can be applied with no modifications also to this framework, however, for the sake of notational simplicity, in the rest of the paper only the case where $\mathcal{S}$ is a partition of $S$ is discussed.} It is assumed that, for each coarsened unit $i$, the region $S_r$ where $i$ is located is known.

To sum up, for all units $i=1,\dots,n$ the values of the dependent variable $y_i$ and the covariates $x_i$ are known. For non-coarsened units $i=1,\dots,p$ the coordinates $c_i\in S$ are known, whereas it is known the coarsening area $S_r$ of each coarsened unit $i=p+1,\dots,n$ such that $c_j\in S_r$. Other missing or unknown information such as the values of parameters and the coordinates of coarsened units about model~\eqref{eq:SLM} should be either learnt (through estimation) or made it non-relevant (through marginalisation).

Before illustrating our proposal for tackling the estimation problem, we introduce the notation that will be used throughout the rest of the paper.

We denote with subscript $P$ and subscript $C$ non-coarsened and coarsened points respectively (that is, points where $\Phi_j=1$ and $\Phi_j=0$ respectively). Conditionally on the random vector $\Phi$, SLM~\eqref{eq:SLM} can be restated as it follows:
\begin{equation}
\label{eq:PSLM}
\begin{bmatrix}
y_P \\
y_C
\end{bmatrix}
=
\rho
\begin{bmatrix}
W_{PP} & W_{PC} \\
W_{CP} & W_{CC}
\end{bmatrix}
\cdot
\begin{bmatrix}
y_P \\
y_C
\end{bmatrix}+
\begin{bmatrix}
X_P \\
X_C
\end{bmatrix}
\beta+
\begin{bmatrix}
\varepsilon_P \\
\varepsilon_C
\end{bmatrix}
\end{equation}
\noindent provided that the original SLM is properly permuted by means of a suitable permutation
matrix $P_\Phi\in\{0,1\}^{n\times n}$, that is:
\begin{align*}
&
\begin{bmatrix}
y_P \\
y_C
\end{bmatrix}
=P_\Phi y\,,
&&
\begin{bmatrix}
W_{PP} & W_{PC} \\
W_{CP} & W_{CC}
\end{bmatrix}
=P_\Phi WP_\Phi\,,
&&
\begin{bmatrix}
X_P \\
X_C
\end{bmatrix}
=P_\Phi X\,,
&&
\begin{bmatrix}
\varepsilon_P \\
\varepsilon_C
\end{bmatrix}
=P_\Phi\varepsilon\,.
\end{align*}

Restatement~\eqref{eq:PSLM} allows to organise observations about coarsened ($C$) and non-coarsened ($P$) points in block matrices.

We also define matrix $A\define I_n-\rho P_\Phi WP_\Phi\in\Real^{n\times n}$, so that:
\[
A=
\begin{bmatrix}
A_{PP} & A_{PC} \\
A_{CP} & A_{CC}
\end{bmatrix}=
\begin{bmatrix}
I_p-\rho W_{PP} & -\rho W_{PC} \\
-\rho W_{CP}    & I_{n-p}-\rho W_{CC}
\end{bmatrix}\,.
\]

Finally, it can be proved (see e.g. \cite{lu2002}) that the following relations hold for the inverse matrix $A\inv$:
\begin{equation}
\label{eq:invAtemp}
A\inv=
\begin{bmatrix}
A_{PP}\inv+A_{PP}\inv A_{PC}\tilde\Xi\inv
A_{CP}A_{PP}\inv &
-A_{PP}\inv A_{PC}\tilde\Xi\inv \\
-\tilde\Xi\inv A_{CP}A_{PP}\inv &
\tilde\Xi\inv
\end{bmatrix}
\end{equation}
\noindent where $\tilde\Xi\define A_{CC}-A_{CP}A_{PP}\inv A_{PC}$ is the Schur complement of $A_{PP}$ and
\begin{align}
A\inv&\define
\begin{bmatrix}
(A\inv)_{PP} & (A\inv)_{PC} \\
(A\inv)_{CP} & (A\inv)_{CC}
\end{bmatrix}=\notag\\
\label{eq:invA}
&=
\begin{bmatrix}
\Xi\inv &
-\Xi\inv A_{PC}A_{CC}\inv \\
-A_{CC}\inv A_{CP}\Xi\inv &
A_{CC}\inv+A_{CC}\inv A_{CP}\Xi\inv
A_{PC}A_{CC}\inv
\end{bmatrix}
\end{align}
\noindent where $\Xi\define A_{PP}-A_{PC}A_{CC}\inv A_{CP}$ is the Schur complement of $A_{CC}$ (see e.g. \cite{horn2013}).

\section{Estimation strategy}
\label{sec:estimation}

\subsection{Model fitting}

Equation~\eqref{eq:invAtemp} allows to restate the reduced form of model~\eqref{eq:PSLM} as follows:
\begin{equation}
\label{eq:SLMP}
y_P
=\rho W_{PP}y_P+X_P\beta+\varepsilon_P
+A_{PC}\Xi\inv
\left[A_{CP}A_{PP}\inv\,(X_P\beta+\varepsilon_P)
-(X_C\beta+\varepsilon_C)\right]\,.
\end{equation}

Left-hand side term of Equation~\eqref{eq:SLMP} together with the first three terms of the right-hand side perfectly describe a SLM amongst correctly geo-referenced points, sharing the same parameters of the complete model~\eqref{eq:SLM}. Unfortunately, the last term on the right-hand side makes things more complicated.

The fourth term on the right-hand side of Equation~\eqref{eq:SLMP} proves that, in general, any subset of observations of a SLM does not follow a SLM. Indeed, it makes the estimation process of a SLM with coarsened points particularly tricky since Equation~\eqref{eq:SLMP} includes blocks of matrix $A$ which depend on (unknown) coordinates of coarsened points.

As previously stated, the estimation strategy proposed in this paper relies on a double marginalisation of SLM~\eqref{eq:SLM}. In particular, the former marginalisation should be made with respect to $y_P$, thus concentrating the information about coarsened points into a lower dimensional space. A similar approach to the marginalisation of the SLM has alredy proved to be successful in the context of variance estimation in 2-dimensional systematic sampling (see \cite{espa2017}). The latter marginalisation should instead be made with respect to the point process of non-coarsened points $Z_P$, so as to include direct and indirect effects of positional errors in the (marginal) probability distribution of $y_P$.

The first marginalisation can be derived in closed form from the inverse formula~\eqref{eq:invA} and equals
\[
y_P
=\Xi\inv X_P\beta+\Xi\inv\varepsilon_P
-\Xi\inv A_{PC}A_{CC}\inv(X_C\beta+\varepsilon_C)\,,
\]
\noindent which is a restatement of Equation~\eqref{eq:SLMP} and implies that:
\begin{subequations}
\label{eq:MomSLMP}
\begin{align}
\E(y_P|Z,\Phi)
&=\Xi\inv X_P\beta+\rho\,\Xi\inv W_{PC}A_{CC}\inv X_C\beta\,,\\
\cov(y_P|Z,\Phi)
&=\sigma^2\,\Xi\inv(I_p+\rho^2\,W_{PC}(A_{CC}\T A_{CC})\inv W_{PC}\T)(\Xi\inv)\T\,.
\end{align}
\end{subequations}

On the other hand, the second marginalisation requires the intensity function $\lambda$ to be estimated, so as to characterise the spatial point process $\set{Z(s,\omega)\colon s\in S}$ and, in turn, the probabilistic law of the spatial weight matrix $W$ under coarsened geocoding.

According to \cite{zimmerman2008}, for any $s\in S$, the intensity function of a spatial point pattern affected by incomplete geocoding can be estimated as follows:
\begin{equation}
\label{eq:lambdaZim}
\hat\lambda(s)=\sum_{i=1}^n[\hat\phi(z_i)]^{-1}K_h(s-z_i),
\end{equation}
where $K$ is some kernel function with bandwidth $h$, $z_i$ is an observed unit's point location, and $\hat\phi$ is an estimate of the geocoding propensity function $\phi\colon S\to(0,1]$ \cite{zimmerman2008}.

The geocoding propensity function $\phi$ can be estimated in various ways, according to the available information about the coarsening process. In this paper, the values of the coarsening probabilities in~\eqref{eq:coarseningprocess} are assumed to be such that $p_j=\phi(z_j)$, given the coordinate $z_j\in S$ of the unit $j$. It follows that:
\begin{equation}
\label{eq:phiZim}
\hat\phi(s)=
\frac{\sum_{r=1}^R\sum_{j=1}^n\Phi_j\ind{z_j\in S_r}\ind{s\in S_r}}
{\sum_{r=1}^R\sum_{j=1}^n\ind{z_j\in S_r}\ind{s\in S_r}}\,,
\end{equation}
so that $\hat\phi$ is constant over each region $S_r\in\mathcal{S}$ and equals the proportion of non-coarsened points in $S_r$.

As stated in \cite{zimmerman2008}, function~\eqref{eq:lambdaZim} can be estimated via a weighted kernel intensity estimator \cite{diggle1985}.

To sum up, the solution we propose in this paper consists in four steps:
\begin{enumerate}
\item the intensity function of the coarsened point process $Z$ is estimated according to \cite{zimmerman2008} through estimators~\eqref{eq:lambdaZim} and~\eqref{eq:phiZim};
\item the likelihood of SLM~\eqref{eq:SLM} marginalised with respect to $y_P$ is derived from~\eqref{eq:MomSLMP}; we denote that likelihood function as $\Lik(\rho,\beta,\sigma^2|y,X,Z,\Phi)$;
\item the likelihood $\Lik(\rho,\beta,\sigma^2|y,X,Z,\Phi)$ is marginalised with respect to $Z_P$, that is:
\begin{equation}
\label{eq:intlogLik}
\Lik(\rho,\beta,\sigma^2|y,X,Z_P,\Phi)=\int_{S^{n-p}}\Lik(\rho,\beta,\sigma^2|y,X,Z_P,z_C,\Phi) \,\hat\varrho(z_C|Z_P)\,\diff z_C
\end{equation}
where $\hat\varrho\colon S^{n-p}\to\Real^+$ is the conditional probability density function of $Z_C|Z_P$ implied by the estimated intensity function $\hat\lambda$;
\item marginal likelihood $\Lik(\rho,\beta,\sigma^2|y,X,Z_P,\Phi)$ is maximised with respect to $\rho$, $\beta$ and $\sigma^2$.
\end{enumerate}

As anticipated, marginalisation~\eqref{eq:intlogLik} has to be performed numerically since it seems
impossible to compute it analytically. Anyway, two issues may make the outlined method computationally unfeasible.

Firstly, the high-dimensional integration space in~\eqref{eq:intlogLik} may subsantially deteriorate the performances of Monte Carlo integration methods.

Secondly, the need to evaluate integral~\eqref{eq:intlogLik} at every step of the optimisation procedure dramatically exacerbates the problem outlined in the previous point.

In order to overcome both problems (and the second in particular), we rely on the cross-entropy algorithm for the optimisation of noisy functions \cite{rubinstein2004}, which iteratively marginalises and optimises the likelihood function $\Lik(\rho,\beta,\sigma^2|y,X,Z_P,Z_C,\Phi)$ at the same time. Results of Monte Carlo simulations discussed in the next section have been performed adopting the same parameters and instrumental distributions of the cross-entropy algorithm as in \cite{bee2017}, where the method have been applied to maximum likelihood estimation of generalised linear multilevel models (the only exception is in the number $N$ of draws, as it will be clarified later).

\subsection{Theoretical properties and generalisations}

As stated in the introduction, this paper aims at proposing an estimation method for spatial models \emph{\`a la} Cliff-Ord \cite{cliff1969} where a portion of data is affected by coarsening, thus the primarily interest is devoted to the parameters of that model, and to the other measures of covariates' effects (like, e.g. direct, indirect and total impacts, which will be discussed in Section~\ref{sec:impacts}).

However, the theoretical properties of the proposed estimation method cannot be easily derived, considering the composite nature of the model, which, in fact, consists of three elements: the point process, the coarsening process, and the spatial model (the SLM in this case). The properties of the estimators of the model's parameters clearly depend on the statistical properties of both the geocoding propensity function estimator~\eqref{eq:phiZim} and the intensity function estimator~\eqref{eq:lambdaZim}. The former is basically a frequency estimator which may be interpreted as an estimator of the mean value of the coarsening intensity function over each coarsening region.  Asymptotic properties of estimator~\eqref{eq:lambdaZim} along with estimator~\eqref{eq:phiZim} are discussed in \cite{zimmerman2008}. Finally, the asymptotics of maximum likelihood estimators of the SLM for non-coarsened data are analysed in depth in \cite{lee2004}.

Since the estimation method proposed in this paper relies on a marginalisation of the full likelihood function of the marked point process which describe all three random processes, the major concern for consistency of the estimators is represented by misspecification problems in the model.

Although that issue is clearly important, it is worth stressing that the modelling approach proposed in this paper can be easily adapted or generalised to other coarsening mechanisms, point patterns, or stochastic spatial processes, as it is only required that the model can be identified and its likelihood marginalised. In general, if the estimators adopted for each component of the model (point process, coarsening process, and spatial process) are singularly consistent, the marginalisation preserves such property for the parameter estimates once the effects of coarsening are considered.

\subsection{Impact estimators}
\label{sec:impacts}

According to \cite{lesage2009}, the effects of covariates on the dependent variable of a SLM do not solely depend on regression coefficients $\beta$, as the spatially-lagged dependent variable induces an indirect effect resulting from the autoregressive parameter $\rho$ and the spatial weight matrix $W$. It follows that the overall impact of a regressor on the value of the dependent variable can be decomposed in a direct and an indirect impact, which, however, it is not constant amongst all units. For these reasons, averages of total ($T(\beta)$), direct ($D(\beta$)), and indirect ($M(\beta)$) impacts are usually computed \cite{lesage2009}:
\begin{subequations}
\label{eq:impacts}
\begin{align}
T(\beta)=n^{-1}\,\iota_n\T(I-\rho W)\inv\iota_n\beta\,,\\
D(\beta)=n^{-1}\,\trace(I-\rho W)\inv\beta\,,\\
M(\beta)=T(\beta)-D(\beta)\,.
\end{align}
\end{subequations}

According to the model we have described in Section~\ref{sec:model}, some elements of the spatial weight matrix $W$ are not known when geocoding is not complete. It follows that impacts should be estimated via Monte Carlo simulations where the weight matrices are generated from realisations of point process $Z$ with estimated intensity function $\hat\lambda$. Thus, the Monte Carlo estimators of the impact measures~\eqref{eq:impacts} can be defined as follows:
\begin{align*}
&\widehat{(A^{-1})}=\frac{1}{N}\sum_{k=1}^N(I-\hat\rho W_k)^{-1}\,,
&&\hat{T}(\hat\beta)=n^{-1}\,\iota_n\T\widehat{(A^{-1})}\iota_n\hat\beta\,,\\
&\hat{D}(\hat\beta)=n^{-1}\,\trace\widehat{(A^{-1})}\hat\beta\,,
&&\hat{M}(\hat\beta)=\hat{T}(\hat\beta)-\hat{D}(\hat\beta)\,.
\end{align*}
Since Monte Carlo estimation of matrix $\widehat{(A^{-1})}$ may be computationally demanding because of the inversions of the weight matrices $W_k$, a truncated geometric series of $(I-\hat\rho W_k)\inv$ may reduce substantially the computational burden of the simulation:
\[
\widehat{(A^{-1})}=\frac{1}{N}\sum_{k=1}^N\sum_{h=0}^m\hat\rho^hW_k^h\,.
\]
where $m$ represents the truncation point.

\section{Monte Carlo simulations}
\label{sec:simulations}

The performances of the proposed estimation approach in finite samples have been studied by means of Monte Carlo simulations. The complication of both the modelling setting and estimation method considerably widens the
variety of scenarios which should be considered for studying the estimators' properties in finite samples.

\begin{figure}[t]
\centering
\caption{Intensity function $\lambda$ used for generating the point process (left) and the realisation of the process for $n=250$ with hexagonal partition of the space (right).}
\includegraphics[scale=0.67]{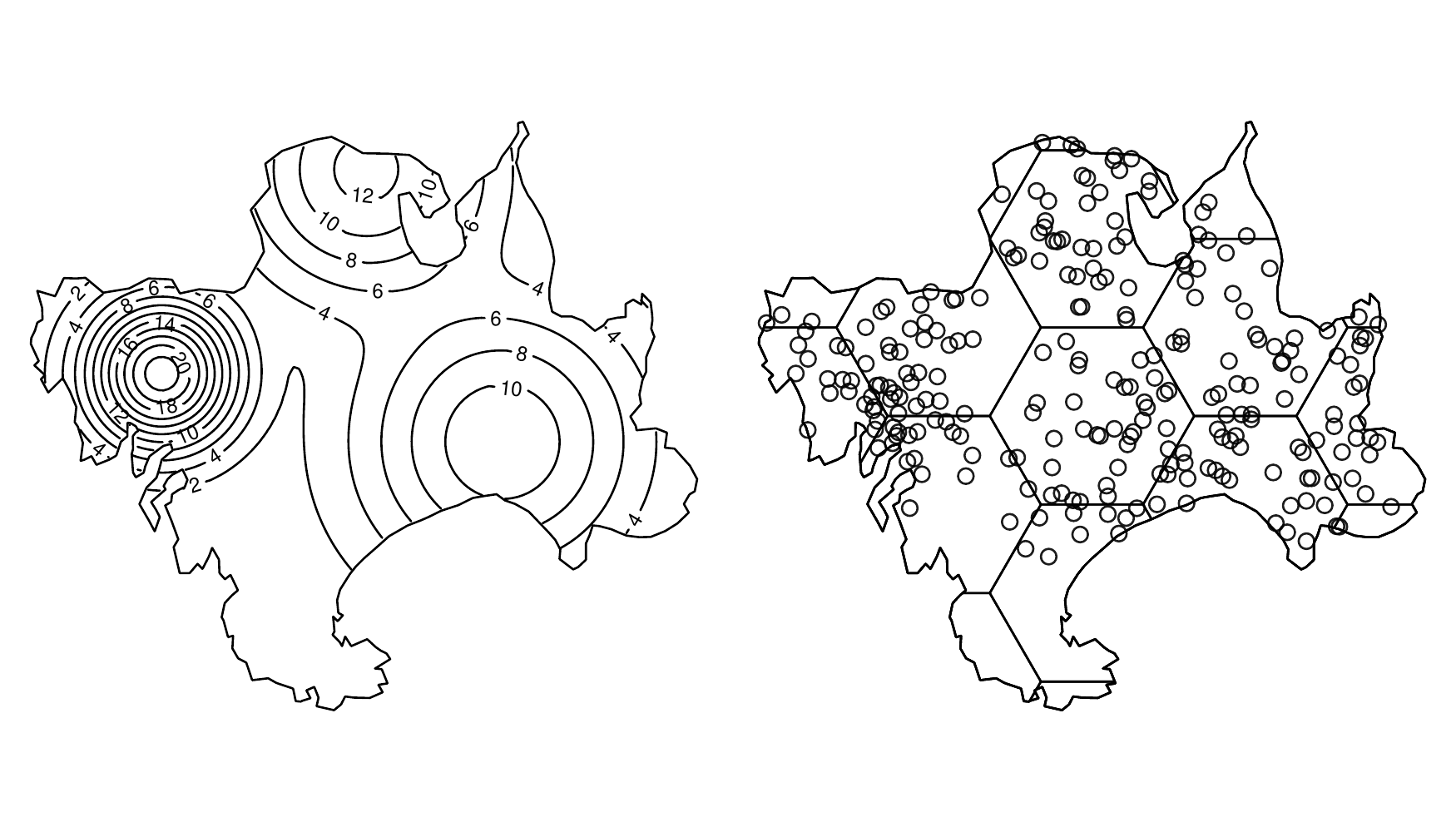}
\label{fig:ppp}
\end{figure}

In this section eight different scenarios are considered:
\begin{enumerate}[(A)]
\item a point pattern with $n=250$ points is generated over an irregular area $S$ according to an inhomogeneous Poisson process with the intensity function $\lambda$ represented in Figure~\ref{fig:ppp}. The surface $S$ is partitioned into $R=17$ hexagonal regions of equal size excepting for border zones (see Figure~\ref{fig:ppp}). The SLM includes two regressors (generated as realisations of a standard normal distribution) and a constant term, so that $X\in\Real^{n\times 3}$. The parameters of the SLM are $\rho=0.5$, $\beta=[1, 1, -1]\T$, $\sigma^2=1$, whereas the spatial weight matrix $W$ is computed according to~\eqref{eq:wfun}, and $\kappa(x)=\ind{x\leq0.5}$ (note that sides of hexagons measure $1.5$).
Each unit of the point pattern is independently coarsened with probability $0.4$. Simulations are based on $N=300$ replications, each of which share the same point pattern and design matrix $X$;
\item the same simulation settings as in point~(A), except that $\rho=0.3$;
\item the same simulation settings as in point~(A), except that $\rho=0.7$;
\item the same simulation settings as in point~(A), except that $\sigma^2=2$;
\item the same simulation settings as in point~(A), except that $n=500$ and $\kappa(x)=\ind{x\leq\sqrt{1/8}}$. Function $\kappa$ has been redefined so that the average neighbourhood area per unit is the same as in case~(A);
\item the same simulation settings as in point~(A), except that $\phi(s)\propto 0.8\,\lambda(s)$. Function $\phi$ is set so that the  coarsening probability ranges between $0.2$ and $0.75$, whereas its average equals $0.4$, in line with all the other simulation scenarios;
\item the same simulation settings as in point~(A), except that $\phi(s)\propto -0.8\,\lambda(s)$. Function $\phi$ is set so that the  coarsening probability ranges between $0.04$ and $0.60$, whereas its average equals $0.4$, in line with all the other simulation scenarios.
\item the same simulation settings as in point~(A), except that the sides of hexagons measure 1, thus the number of regions is $R=29$;
\end{enumerate}

For each scenario five estimation methods are considered:
\begin{itemize}
\item the maximum likelihood estimator based on a dataset where location of all units are known, and there is no coarsening. Hereinafter this estimator is referred to as NCM, which stands for \emph{non-coarsened model};
\item the proposed estimator based on double marginalisation (hereinafter DME);
\item the maximum likelihood estimator of the SLM based only on non-coarsened units (hereinafter REM). In this case the weight matrix is computed using the same $\kappa$ function as the data generating process, but no standardisation is performed;
\item the maximum likelihood estimator of the SLM based only on non-coarsened units. Unlike the previous case, the spatial weight matrix is row-standardised (hereinafter SREM);
\item the maximum likelihood estimator of the SLM based on all points. Location of coarsened points is imputed to the centroids of regions where points are located, and a row-standardised weight matrix is derived according to the same $\kappa$ function as the data generating process. Hereinafter this method is referred to as CIP, which stands for \emph{centroid imputed position}.
\end{itemize}

Results of simulations are summarized in terms of relative root mean squared error (RMSE) and relative bias in Tables~\ref{tab:simulABCD} and~\ref{tab:simulEFG}. Tables~\ref{tab:simulABCD} and~\ref{tab:simulEFG} only report impacts estimates about the first regressor, since estimates on other regressor impacts are similar.

\begin{table}[t!]
\centering
\scriptsize
\caption{Relative root mean squared error and relative bias (in parenthesis) of parameter and impact estimators for scenarios A, B, C, D of Monte Carlo simulations (see Section~\ref{sec:simulations}) for various estimation methods. Direct (D), indirect (M) and total (T) impact estimates refer to the second regressor (whose coefficient is $\beta_1$). All values are multiplied by 100.}
\label{tab:simulABCD}
\begin{tabular}{@{} l *{8}{d{3.2}}@{}} 
\toprule
{Method} & \rho & \beta_0 & \beta_1 & \beta_2 & \sigma & D(\beta_1) & M(\beta_1) & T(\beta_1) \\
\midrule
\multicolumn{9}{c}{Scenario A} \\
\midrule
\multirow{2}{*}{NCM} & 4.78 & 10.40 & 2.29 & 3.85 & 4.20 & 2.43 & 9.57 & 5.22 \\
 & (-0.40) & (-0.70) & (-0.11) & (-0.15) & (-0.59) & (-0.14) & (-0.46) & (-0.28) \\
\multirow{2}{*}{DME} & 23.74 & 24.58 & 3.99 & 6.16 & 25.41 & 4.59 & 37.50 & 18.81 \\
 &  (-22.25) & (-17.98) & (0.68) & (0.28) & (24.09) & (-2.57) & (-36.00) & (-17.89) \\
\multirow{2}{*}{SREM}  & 29.67 & 30.19 & 3.80 & 5.80 & 25.70 & 3.97 & 48.34 & 23.10 \\
 & (-28.50) & (-24.85) & (0.74) & (0.82) & (24.23) & (-1.33) & (-47.44) & -22.46 \\
\multirow{2}{*}{CIP}  & 33.01 & 27.50 & 3.14 & 4.57 & 44.09 & 3.77 & 47.87 & 23.46 \\
 & (-32.11) & (-23.75) & (1.48) & (1.36) & (43.36) & (-2.61) & (-47.06) & (-22.98) \\
\multirow{2}{*}{REM}  & 83.95 & 39.36 & 4.13 & 5.88 & 40.72 & 4.11 & 49.03 & 23.78 \\
 & (-83.89) & (-34.71) & (2.00) & (-0.02) & (39.54) & (-1.67) & (-43.44) & (-20.82) \\
\addlinespace
\multicolumn{9}{c}{Scenario B} \\
\midrule
\multirow{2}{*}{NCM} & 9.84 & 9.93 & 2.17 & 3.50 & 4.51 & 2.20 & 13.81 & 4.63\\
 & (-0.19) & (-0.63) & (-0.05) & (-0.23) & (-0.85) & (-0.03) & (0.22) & (0.04) \\
\multirow{2}{*}{DME} & 29.15 & 17.39 & 3.18 & 5.01 & 9.79 & 3.42 & 36.19 & 11.44\\
 & (-25.81) & (-9.36) & (-0.22) & (0.06) & (7.32) & (-1.27) & (-32.69) & (-10.08) \\
\multirow{2}{*}{SREM}  & 36.51 & 19.10 & 3.00 & 4.84 & 9.71 & 3.19 & 47.20 & 14.26 \\
 & (-34.22) & (-12.70) & (-0.18) & (0.38) & (7.33) & (-0.96) & (-45.38) & (-13.42) \\
\multirow{2}{*}{CIP}  & 38.88 & 17.78 & 2.26 & 3.62 & 14.22 & 2.71 & 46.52 & 14.27 \\
 & (-36.64) & (-13.23) & (-0.19) & (-0.18) & (13.07) & (-1.50) & (-44.43) & (-13.54) \\
\multirow{2}{*}{REM}  & 84.15 & 20.88 & 3.06 & 4.87 & 12.89 & 3.35 & 49.32 & 15.05 \\
 & (-84.05) & (-14.62) & (0.00) & (-0.23) & (10.99) & (-1.16) & (-45.28) & (-13.53) \\
\addlinespace
\multicolumn{9}{c}{Scenario C} \\
\midrule
\multirow{2}{*}{NCM} & 2.42 & 10.99 & 2.21 & 3.76 & 4.92 & 2.36 & 7.79 & 5.54\\
 & (-0.34) & (-0.93) & (-0.20) & (-0.13) & (-0.86) & (-0.33) & (-0.91) & (-0.70) \\
\multirow{2}{*}{DME} & 17.58 & 42.81 & 4.86 & 8.64 & 58.44 & 7.08 & 43.65 & 30.03 \\
 & (-16.66) & (-33.12) & (2.04) & (1.03) & (56.86) & (-5.79) & (-42.98) & (-29.48) \\
\multirow{2}{*}{SREM}  & 22.30 & 51.97 & 5.29 & 7.21 & 59.68 & 4.53 & 51.00 & 33.02 \\
 & (-21.60) & (-44.62) & (3.24) & (2.08) & (57.89) & (-0.96) & (-50.39) & (-32.45) \\
\multirow{2}{*}{CIP}  & 28.41 & 49.80 & 6.70 & 7.43 & 110.29 & 5.39 & 53.90 & 36.01 \\
 & (-27.92) & (-45.83) & (5.81) & (5.18) & (109.49) & (-4.42) & (-53.50) & (-35.69) \\
\multirow{2}{*}{REM}  & 84.69 & 86.62 & 8.58 & 8.46 & 109.38 & 5.54 & 48.59 & 32.33 \\
 & (-84.66) & (-81.77) & (7.35) & (2.63) & (108.13) & (-2.27) & (-40.87) & (-26.86) \\
\addlinespace
\multicolumn{9}{c}{Scenario D} \\
\midrule
\multirow{2}{*}{NCM}
& 6.35 & 14.79 & 3.30 & 4.97 & 4.87 & 3.30 & 12.12 & 6.51 \\
& (-0.76) & (0.69) & (0.37) & (0.54) & (-0.80) & (0.30) & (-0.42) & (-0.03) \\
\multirow{2}{*}{DME}
& 24.23 & 27.68 & 5.00 & 7.58 & 15.30 & 5.13 & 37.42 & 18.65 \\
& (-21.96) & (-16.14) & (1.15) & (1.23) & (13.30) & (-2.06) & (-35.01) & (-17.16) \\
\multirow{2}{*}{SREM}
& 30.77 & 31.76 & 4.79 & 7.48 & 16.03 & 4.60 & 48.73 & 23.08 \\
& (-29.17) & (-22.53) & (1.40) & (1.84) & (14.30) & (-0.77) & (-47.46) & (-22.17) \\
\multirow{2}{*}{CIP}
& 34.08 & 29.41 & 4.20 & 6.01 & 27.21 & 4.12 & 48.42 & 23.50 \\
& (-32.67) & (-22.32) & (2.03) & (2.09) & (26.27) & (-2.09) & (-47.14) & (-22.74) \\
\multirow{2}{*}{REM}
& 84.20 & 39.79 & 5.34 & 7.35 & 25.13 & 4.86 & 48.32 & 23.31 \\
& (-84.14) & (-31.45) & (2.56) & (0.82) & (23.89) & (-1.18) & (-42.58) & (-20.15) \\
\bottomrule
\end{tabular}
\end{table}

\begin{table}[t!]
\centering
\scriptsize
\caption{Relative root mean squared error and relative bias (in parenthesis) of parameter and impact estimators for scenarios E, F, G, H of Monte Carlo simulations (see Section~\ref{sec:simulations}) for various estimation methods. Direct (D), indirect (M) and total (T) impact estimates refer to the second regressor (whose coefficient is $\beta_1$). All values are multiplied by 100.}
\label{tab:simulEFG}
\begin{tabular}{@{} l *{8}{d{3.2}}@{}} 
\toprule
{Method} & \rho & \beta_0 & \beta_1 & \beta_2 & \sigma & D(\beta_1) & M(\beta_1) & T(\beta_1) \\
\midrule
\multicolumn{9}{c}{Scenario E} \\
\midrule
\multirow{2}{*}{NCM}
& 3.42 & 7.01 & 1.47 & 2.36 & 3.44 & 1.52 & 6.68 & 3.52 \\
& (-0.17) & (-0.20) & (0.03) & (-0.07) & (-0.70) & (0.02) & (-0.08) & (-0.02) \\
\multirow{2}{*}{DME}
& 26.50 & 24.73 & 3.36 & 5.17 & 25.81 & 3.46 & 40.36 & 19.73 \\
& (-25.62) & (-20.79) & (1.87) & (2.65) & (24.89) & (-2.09) & (-39.42) & (-19.14) \\
\multirow{2}{*}{SREM}
& 29.77 & 26.93 & 3.17 & 4.97 & 25.99 & 2.68 & 47.67 & 22.28 \\
& (-29.01) & (-23.84) & (1.94) & (2.86) & (25.02) & (-0.65) & (-47.01) & (-21.83) \\
\multirow{2}{*}{CIP}
& 33.62 & 24.32 & 2.96 & 5.52 & 40.96 & 2.91 & 48.02 & 23.23 \\
& (-33.00) & (-21.96) & (2.34) & (4.48) & (40.44) & (-2.26) & (-47.43) & (-22.90) \\
\multirow{2}{*}{REM}
& 85.04 & 34.12 & 4.10 & 7.02 & 41.29 & 2.95 & 37.40 & 17.85 \\
& (-85.02) & (-31.62) & (3.18) & (5.72) & (40.67) & (-0.21) & (-26.80) & (-12.35) \\
\addlinespace
\multicolumn{9}{c}{Scenario F} \\
\midrule
\multirow{2}{*}{NCM}
& 4.67 & 10.23 & 2.14 & 3.72 & 4.36 & 2.18 & 9.09 & 4.82 \\
& (-0.10) & (0.27) & (0.01) & (-0.02) & (-0.86) & (0.03) & (0.22) & (0.12) \\
\multirow{2}{*}{DME}
& 21.86 & 24.29 & 3.66 & 5.96 & 24.63 & 4.08 & 35.26 & 17.55 \\
& (-20.49) & (-17.05) & (0.84) & (0.33) & (23.43) & (-2.12) & (-33.81) & (-16.64) \\
\multirow{2}{*}{SREM}
& 27.07 & 29.30 & 3.56 & 5.75 & 24.73 & 3.50 & 45.52 & 21.30 \\
& (-26.00) & (-23.52) & (1.15) & (0.85) & (23.40) & (-0.34) & (-44.62) & (-20.63) \\
\multirow{2}{*}{CIP}
& 31.00 & 25.80 & 2.95 & 4.30 & 41.10 & 3.57 & 45.71 & 22.40 \\
& (-30.14) & (-21.94) & (1.43) & (1.28) & (40.46) & (-2.40) & (-44.85) & (-21.86) \\
\multirow{2}{*}{REM}
& 80.09 & 33.36 & 3.67 & 5.86 & 37.55 & 3.72 & 46.57 & 22.36 \\
& (-80.05) & (-27.57) & (1.55) & (0.81) & (36.58) & (-1.32) & (-44.73) & (-21.21) \\
\addlinespace
\multicolumn{9}{c}{Scenario G} \\
\midrule
\multirow{2}{*}{NCM}
& 5.04 & 10.24 & 2.11 & 3.85 & 4.58 & 2.19 & 9.70 & 5.14 \\
& (-0.48) & (-1.40) & (0.05) & (-0.19) & (-1.01) & (0.01) & (-0.42) & (-0.19) \\
\multirow{2}{*}{DME}
& 24.43 & 27.55 & 3.63 & 6.04 & 24.82 & 4.26 & 38.00 & 19.01 \\
& (-22.95) & (-21.33) & (0.90) & (0.13) & (23.33) & (-2.63) & (-36.53) & (-18.17) \\
\multirow{2}{*}{SREM}
& 31.01 & 33.33 & 3.50 & 5.69 & 25.33 & 3.56 & 49.13 & 23.46 \\
& (-29.85) & (-28.52) & (1.26) & (0.54) & (23.89) & (-1.41) & (-48.23) & (-22.86) \\
\multirow{2}{*}{CIP}
& 32.87 & 27.86 & 3.04 & 4.67 & 44.74 & 3.34 & 47.38 & 23.09 \\
& (-31.94) & (-24.23) & (1.86) & (1.34) & (44.01) & (-2.34) & (-46.54) & (-22.59) \\
\multirow{2}{*}{REM}
& 89.09 & 50.58 & 4.86 & 5.73 & 43.11 & 4.09 & 54.25 & 26.19 \\
& (-89.06) & (-46.77) & (3.27) & (-1.09) & (42.11) & (-1.71) & (-48.39) & (-23.10) \\
\addlinespace
\multicolumn{9}{c}{Scenario H} \\
\midrule
\multirow{2}{*}{NCM}
& 4.84 & 9.44 & 2.01 & 3.40 & 4.93 & 2.16 & 9.57 & 5.13 \\
& (-0.27) & (-0.47) & (0.01) & (-0.23) & (-0.60) & (0.00) & (-0.07) & (-0.03) \\
\multirow{2}{*}{DME}
& 20.28 & 22.43 & 3.52 & 6.00 & 24.59 & 4.16 & 33.16 & 16.73 \\
& (-18.59) & (-15.55) & (0.47) & (0.14) & (22.91) & (-2.41) & (-31.33) & (-15.66) \\
\multirow{2}{*}{SREM}
& 29.18 & 29.64 & 3.50 & 5.48 & 25.88 & 3.48 & 47.55 & 22.48 \\
& (-28.07) & (-24.61) & (1.18) & (0.57) & (24.12) & (-0.88) & (-46.65) & (-21.85) \\
\multirow{2}{*}{CIP}
& 27.94 & 22.79 & 3.12 & 4.21 & 40.35 & 3.27 & 41.89 & 20.44 \\
& (-26.98) & (-19.15) & (1.76) & (1.55) & (39.58) & (-2.05) & (-40.91) & (-19.86) \\
\multirow{2}{*}{REM}
& 84.05 & 38.60 & 4.28 & 5.87 & 40.73 & 4.02 & 50.74 & 24.49 \\
&(-84.00) & (-34.50) & (2.54) & (-0.43) & (39.48) & (-1.16) & (-42.00) & (-19.87) \\
\bottomrule
\end{tabular}
\end{table}

Obviously, in all scenarios, the NCM estimator is the best performer for all parameters both in terms of bias and RMSE, as it relies on correct positions for all units. For this reason, it is not commented in the following.

Estimates in Tables~\ref{tab:simulABCD} and~\ref{tab:simulEFG} show two general results which basically hold under all scenarios.

Firstly, the estimates obtained from all estimation methods are rather stable under all simulation settings for most parameters and impacts. The only remarkable exception is represented by the estimates of the error variance, which are rather sensitive with respect to the value of parameter $\rho$ and $\sigma^2$.

Secondly, the rank of estimation methods in terms of both bias and RMSE is basically the same whatever the scenario we consider, although some differences emerge amongst parameters.

If covariate coefficients are considered (that is $\beta_0$, $\beta_1$, $\beta_2$), DME estimator is the best performer in terms of relative bias. On the other hand, the CIP estimator exhibits the smallest RMSE, followed by the SREM estimator, whereas larger RMSE result from DME and REM estimator. Anyway, both in case of bias and RMSE, differences amongst estimators are rather small if we consider covariates coefficients $\beta_1$ and $\beta_2$, whereas larger variability emerges for $\beta_0$.

Things change if the autoregressive parameter $\rho$ is considered. In this case, the DME clearly outperforms all other estimators both in terms of bias and RMSE in all considered scenarios, whereas the second-best estimator is SREM estimator followed by CIP and REM estimators. Unlike regressors coefficients, differences amongst estimation methods are large in terms of bias and RMSE.

If error dispersion parameter $\sigma$ is considered, the four estimation methods for coarsened data can be gathered into two groups. The former includes the best performers which are DME and SREM, the latter consists in CIP and REM estimators, which almost double the relative bias and the relative RMSE of estimators in the other group. It is interesting to note that estimators of each group exhibit very similar relative bias and relative RMSE.

The performances of estimators on assessing impacts of covariates clearly reflect the statistical performances on parameters $\rho$, $\beta_1$, and $\beta_2$. Thus CIP, REM, and SREM estimators perform well in estimating the direct impact, whereas the DME definitely outperforms the others when indirect impact is estimated. The efficiency of DME on indirect impact estimation is large enough to make DME the most efficient estimator also for the total impact. Analogous results hold also in terms of bias.

Although relative performances of estimators are pretty stable amongst scenarios considered in the simulations, it is worth stressing some stylised facts which emerged from simulations and are in line with the behaviour which may be expected.

Firstly, the relative bias and the relative RMSE of the estimators of the autoregressive parameter $\rho$ and the error dispersion parameter $\sigma$ are associated with $\rho$ itself. In particular, the larger is $\rho$, the smaller will be the relative bias and the relative RMSE of estimators for $\rho$ and $\sigma$. This property seems to hold also for the other parameters ($\beta_0$, $\beta_1$, $\beta_2$), however the magnitude of the effect is not particularly wide.

The bias and the RMSE of impact estimates are related to the value of $\rho$ too, as they depend on bias and RMSE  of estimates on parameters $\rho$, $\beta_1$, $\beta_2$, thus, the higher the value of $\rho$, the higher is the efficiency of the considered estimators.

Secondly, scenario (D) shows that, as expected, an increase in the error variance with respect to scenario (A), leads to a loss in efficiency of all estimation methods. On the other hand, if the size of the regions is reduced, the effect of coarsening are more limited, and this turns in to an increase of estimators efficiency and a decrease of biases, as the comparison of results from scenario (A) and (H) makes it apparent.

Thirdly, if scenarios (A), (F), and (G) are compared, no clear pattern emerges, although it seems that RMSE tends to slightly increase as we move from scenario (F) to (A), and from (A) to (G), suggesting that better estimates can be obtained if coarsening is more frequent in areas where the intensity of the point process is higher -- scenario (F) --, whereas the opposite is true if the intensity of the point process and the coarsening probability are inversely related.

\section{Conclusions}
\label{sec:conclusions}

The estimation method proposed in this paper for tackling the problem of incompletely geocoded data is based on a modelling approach which integrates the point process, the coarsening process and the spatial process through a marked point process model whose likelihood function is then marginalised twice so as to clean out the effects of coarsening.

Monte Carlo simulations for the spatial lag model have shown that the proposed method is basically equivalent to other methods in terms of bias and RMSE in the estimation of regressor coefficients, whereas it returns more efficient and less biased estimates for the spatial autoregressive parameter, the error variance, indirect impacts, and total impacts. Gains in efficiency and biasedness are substantial and they clearly emerges under the various simulation settings. The proposed methodology can be generalised in various directions to account for other forms of data incompleteness typically emerging when analysing large spatial datasets related to individual economic agents.

\printbibliography

\end{document}